\documentclass[twocolumn,showpacs,preprintnumbers,amsmath,amssymb]{revtex4}
\usepackage{graphicx}
\usepackage{dcolumn}
\usepackage{bm}
\usepackage{multirow}

\newcommand{\Xstate}{X~$^1\Sigma^+$}
\newcommand{\astate}{a~$^3\Sigma^+$}

\begin{document}

\title{Feshbach resonances in an ultracold $^7$Li and $^{87}$Rb mixture}

\author{C. Marzok, B. Deh, C. Zimmermann, and Ph.\,W. Courteille}
\affiliation{Physikalisches Institut, Eberhard-Karls-Universit\"at
T\"ubingen, Auf der Morgenstelle 14, D-72076 T\"ubingen,
Germany}
\author{E. Tiemann}
\affiliation{Institut f\"ur Quantumoptik, Leibniz Universit\"at
Hannover,\\Welfengarten 1, D-30167 Hannover, Germany}
\author{Y.\,V. Vanne} \author{A. Saenz}
\affiliation{AG Moderne Optik, Institut f\"ur Physik, Humboldt-Universit\"at zu Berlin, Hausvogteiplatz 5-7, D-10117
Berlin, Germany}

\date{\today}

\begin{abstract}
We report on the observation of five Feshbach resonances in
collisions between ultracold $^7$Li and $^{87}$Rb atoms in the
absolute ground state mixture where both species are in their
$|f,m_f\rangle=|1,1\rangle$ hyperfine states. The resonances
appear as trap losses for the $^7$Li cloud induced by inelastic
heteronuclear three-body collisions. The magnetic field values where
they occur are important quantities for an accurate determination
of the interspecies interaction potentials. Results of coupled channels
calculations based on the observed resonances are presented and refined
potential parameters are given. A very broad Feshbach
resonance centered around 649\,G should allow for fine
tuning of the interaction strength in future experiments.
\end{abstract}


\maketitle

Feshbach spectroscopy has enabled physicists to gather high
precision data on interaction potentials that made it possible to
significantly increase the knowledge of the potential curves, since the location of
Feshbach resonances is very sensitive to the shape of the
long-range part of the interatomic potentials. Their experimental
observation therefore sets tight constraints on them and fixes the
scattering lengths. This works also for mixtures of different
atomic species, where Feshbach resonances have been found in a
range of alkali mixtures. To date these include Fermi-Bose
mixtures such as $^6$Li+$^{23}$Na \cite{Stan2004} and
$^{40}$K+$^{87}$Rb \cite{Inouye2004,Klempt2006}, Bose-Bose mixtures
like $^{85}$Rb+$^{87}$Rb \cite{Papp2006} and $^{39}$K+$^{87}$Rb
\cite{Simoni2008} as well as very recently Fermi-Fermi mixtures
like $^6$Li+$^{40}$K \cite{Wille2008}.

Until a few months ago, in the absence of spectroscopic data for
the Li-Rb mixtures, the interaction potentials were unknown to an
extent that not even the signs of the background scattering
lengths were known \cite{Silber2005,Marzok2007}. Recently, this
has changed with the observation of heteronuclear Feshbach
resonances in the Fermi-Bose mixture of $^6$Li and $^{87}$Rb atoms
in the absolute ground state \cite{Deh2008}. In
parallel to our own runing analysis, the group of K. W. Madison (UBC, Vancouver) has been able
to determine the closed channels underlying the resonances by means of theoretical coupled channel
calculations based on \textit{ab initio} interaction potentials \cite{Li2008}.
But they constructed artificial potentials, with which
extrapolation to other isotope combinations by mass scaling is
normally not reliable. At the same time we did a similar analysis
applying potential curves of the singlet and triplet states from
the 2s+5s atomic asymptote of Li+Rb, which were obtained by
Fourier transform spectroscopy \cite{pashov08}, with which
predictions are meaningful for searches of Feshbach resonances for other isotopic
combinations.

Furthermore, for the $^7$Li+$^{87}$Rb system, there is a discrepancy
between the experimentally determined absolute value of triplet
scattering length \cite{Marzok2007} and the calculations described
in \cite{Li2008}. Now, with the observation of heteronuclear
Feshbach resonances in the purely bosonic system of
$^7$Li+$^{87}$Rb, new data becomes available that allows for precise
calculations of the Bose-Bose interaction potentials by also
incorporating the new spectroscopic results by Pashov et al \cite{pashov08}.

Accurate knowledge of potential curves is crucial for the
efficient production of ultracold polar molecules. Here, the Li-Rb
system is of particular interest, as the huge difference in atomic number Z
leads to large electric dipole moments predicted in \cite{Aymar2005}.
Usually, Feshbach or rf-associated molecules are highly excited
and exhibit no significant dipole moment. Different possible
routes to lower lying vibrational levels include direct population
by means of photoassociation \cite{Schloeder2001}. Huge increases
in the molecule production rates have been predicted for both homonuclear \cite{vanAbeelen1998}
and heteronuclear photoassociation \cite{Pellegrini2008} performed in the vicinity of a Feshbach resonance.
For homonuclear photoassociation this increase has already been shown experimentally \cite{Courteille1998,Junker2008}.

In a Feshbach resonance, the scattering length changes as a
function of magnetic field \cite{Inouye1998}, which allows tuning the strength of
interactions in atomic model systems for many body physics for a wide range of
applications in quantum gas experiments as studied in homonuclear system, but heteronuclear mixtures will open a wide window to different model systems. Further, the recent observation of Efimov resonances in ultracold
collisions of Cs atoms \cite{Kraemer2006} pave the way for
studying higher order bound state systems. Possible observations
of the predicted universal properties are of particular interest
\cite{Braaten2006}. Studies with three identical bosons are
hampered by a large scaling factor for the appearance of these
resonances. The strong mass dependence of the scaling factor in
favor of large mass ratios of the collision partners together with
the newly found very broad Feshbach resonances make the Li+Rb
system an ideal candidate for studying this field.

\bigskip

The experimental sequence for producing ultracold mixtures of
$^7$Li and $^{87}$Rb is similar to that described in
Ref.\,\cite{Marzok2007}. In short, $^7$Li and $^{87}$Rb atoms are
loaded in a magneto-optical trap and are then transferred via several
intermediate magnetic traps into a Ioffe-Pritchard type trap
characterized by the secular frequencies
$\tilde{\omega}/2\pi=(206\times200\times50)^{1/3}\,$Hz (for Rb)
and the magnetic field offset $3.5\,$G. Both species are in their
stretched spin states $|f,m_f\rangle=|2,2\rangle$. The rubidium cloud is
selectively cooled by microwave-induced forced evaporation, whereas the
$^7$Li cloud is cooled sympathetically by interspecies
thermalization to temperatures around $3\,\mu$K. After the cooling
ramp, the magnetic trap is slightly decompressed and moved exactly
into the center of the trap coils. This is important because the
magnetic field gradients due to the small size of the trap coils,
which are also used for the Feshbach search, would otherwise
strongly limit the magnetic field resolution. The atoms are then
transferred into a horizontally crossed dipole trap operating at
1064\,nm. The light is divided into two horizontal laser beams
having powers of 2.9\,W and 3.2\,W, respectively, which intersect
at right angle at the center of the magnetic trap. Both beams are
focused down to $58\,\mu$m resulting in trap depths of $130\,\mu$K
($^{87}$Rb) and $44\,\mu$K ($^7$Li) and trap frequencies of
$\tilde{\omega}_{\mathrm{Rb}}/2\pi=(610\times440\times430)^{1/3}\,$Hz
and
$\tilde{\omega}_{\mathrm{Li}}/2\pi=(1250\times900\times870)^{1/3}\,$Hz.
The mixture thermalizes in the dipole trap at temperatures of
$10\,\mu$K ($^{87}$Rb) and $8\,\mu$K ($^7$Li) with about $5\times10^5$ Rb
atoms and a few times $10^4$ Li atoms. The temperature
difference is due to permanent evaporative cooling of the $^7$Li
cloud in the dipole potential which is shallower for Li than for Rb.

\bigskip

\begin{figure}[ht]
    \centerline{\scalebox{.28}{\includegraphics{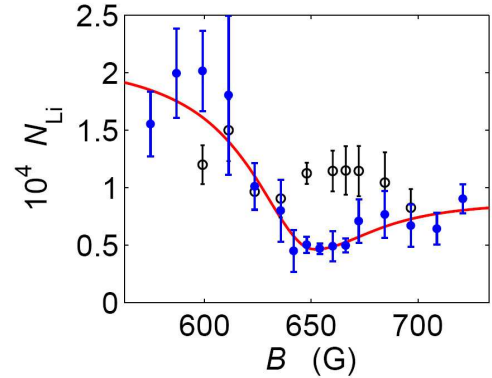}}}
    \caption{(Color online) \label{Fig1} A very broad Feshbach
    resonance has been found located at $B_0=649\,$G. Due to the tight
    confinement in the optical dipole trap, atom losses are observed
    even for fast ramping across the resonance position. The broad
    resonance should allow for precise tuning of the interaction
    strength in future experiments. The magnetic field was held for
    20\,ms at each value. The (black open) circles represent the
    $^7$Li atom number in the absence of $^{87}$Rb while the (blue)
    dots represent $^7$Li with $^{87}$Rb present in the trap. The
    (red) line is a fit based on two Lorentzian curves of same widths
    but different amplitudes and offsets connected at the resonance
    position. It takes atom losses during the fast sweep across the
    resonance when probing the high-field side into account.}
\end{figure}

For the search of Feshbach resonances, the atoms are prepared in their absolute
ground states $|f,m_f\rangle=|1,1\rangle$ by means of rapid
adiabatic passage on their respective hyperfine transition
$|2,2\rangle\rightarrow|1,1\rangle$ at 6.835\,GHz ($^{87}$Rb) and
803\,MHz ($^{7}$Li). Both sweeps are performed in the presence of
a homogeneous 4\,G magnetic field spanning a range of 3\,MHz in
4\,ms. We obtain nearly 100\% transfer efficiency. Residual Rb
$\left|2,2\right>$ is removed by means of a resonant light pulse.
Then the homogeneous magnetic field is quickly increased with fast
ramping speeds between 100 and 600\,G/ms to a variable final value
$B$, where it is either scanned over a limited field interval or
it is held at a fixed value for a certain amount of time. Then the
magnetic field is ramped down again, the dipole trap laser is
turned off, and both clouds are absorption-imaged after a free
expansion time of 300\,$\mu$s for $^7$Li and 5\,ms for $^{87}$Rb.
Atom numbers are determined from these images and measured as a
function of magnetic field. Feshbach resonances are identified by
a dramatic increase in inelastic collision rates showing up in a
reduction of atom numbers. Here, interspecies Feshbach resonances
enhance the three-body loss rate coefficients
$K_{\mathrm{Li,Rb,Rb}}$ and $K_{\mathrm{Li,Li,Rb}}$. Furthermore,
since the Rb atom number exceeds the Li atom number by more than
an order of magnitude, losses are visible specifically on the Li
component. We searched the magnetic field range between 0\,G and
750\,G for increased trap losses. The range is divided into
intervals of 6\,G, and the intervals are swept in subsequent
experimental runs by magnetic field ramps of 200\,ms duration.
Intervals with sudden atom loss are analyzed in more detail.
Several Feshbach resonances are identified at the locations listed
in Table\,\ref{TabI}. For resolving the individual
resonances, different storage times with fixed magnetic fields
(given as $t_{\text{hold}}$, i.e. time on resonance for the individual cases in
Table\,\ref{TabI} ) are used in order to get an optimal
saturation. The profiles are then fitted with Lorentzian functions
to extract the center locations and the widths of the loss
features. For the broad resonance at 649\,G, the fit function is
altered to take atom loss while scanning across the resonance into
account. It consists of two Lorentzians with different offsets and
amplitudes but the same width that are joined together at the
resonance position. In order to resolve this resonance, crossing
over to the high magnetic field side has to be done very quickly.
But still a significant fraction of atoms is lost during the ramp
across the resonance point as shown in Fig.\,\ref{Fig1}. The
asymmetry of the lineshape due to the losses is apparent.

    \begin{table}[h]\tabcolsep2.2mm
        \caption{List of experimentally observed Feshbach resonances in both the Fermi-Bose mixture $^6$Li-$^{87}$Rb
        \cite{Deh2008} and the Bose-Bose mixture $^7$Li-$^{87}$Rb with the resonances positions $B_0$
        and full width at half maximum $\delta B$ of the trap loss
        features for the used magnetic hold time $t_{\mathrm{hold}}$.}\vspace{2mm}
        \renewcommand{\arraystretch}{1.8}
        \begin{tabular}[c]{cccc}\hline\hline
            open channel & $B\,(\text{G})$ & $\delta B\,(\text{G})$& $t_{\mathrm{hold}}$ (ms)  \\\hline
            $^6\text{Li}\,|\frac{1}{2},\frac{1}{2}\rangle\,^{87}\text{Rb}\,|1,1\rangle$ & 882.02 & 1.27 &  75 \\
                                                                  & 1066.92 & 10.62 & 20 \\\hline\hline
            $^7\text{Li}\,|1,1\rangle\,^{87}\text{Rb}\,|1,1\rangle$   & 389.5(2) & 0.9(2) & 30 \\
                                                                  & 447.4(2) & 1.1(2) & 10 \\
                                                                  & 535.4(3) & 1.1(4) & 20 \\
                                                                  & 565(1) & 6(1) & 90     \\
                                                                  & 649(5) & 70(10) & 20 \\\hline
        \end{tabular}
        \label{TabI}
    \end{table}


\bigskip

To model the Feshbach resonances we use the Hamiltonian for the
two electronic states \Xstate and \astate of a pair of atoms $A$
and $B$ with electron and nuclear spin $s$ and $i$, respectively,
but no orbital angular momentum (see for example
Refs.~\cite{mie00,lau02,pas07}):

\begin{eqnarray}
\label{eq:ham}
 H  &=& T_n+U_{\rm X}(R)P_{\rm X} + U_{\rm a}(R)P_{\rm a} \nonumber\\
 & &+a_A\vec{s}_A\vec{i}_A+a_B\vec{s}_B\vec{i}_B\nonumber\\
 & &+[(g_{sA}s_{zA}-g_{iA}i_{zA})+(g_{sB}s_{zB}-g_{iB}i_{zB})]\mu_BB_z\nonumber\\
 & &+\frac{2}{3} \lambda(R)(3S_Z^2-S^2).
\end{eqnarray}

The first line shows the kinetic energy $T_n$ and the potential
energy $U_{\rm X}$ and $U_{\rm a}$ for the internal nuclear motion
of the atom pair; $P_{\rm X}$ and $P_{\rm a}$ are projection
operators on the uncoupled singlet state X and triplet state a,
respectively, for atoms with electron spin 1/2. The second line
shows the hyperfine interaction, determined mainly by the Fermi
contact term, for which we take the magnetic hyperfine parameters
of the atomic ground state of the Li and Rb isotopes from the
report in \cite{ari77}. Feshbach resonances probe mainly the
interaction at the atomic asymptote, thus effects by chemical
bonds in the hyperfine interaction are very small and cannot be
identified in our analysis. The coupling of nucleus $A$ with the
electron spin of atom $B$ is neglected. The third line gives the
Zeeman energy from the electron spin and the nuclear spin by an
external homogeneous magnetic field $B$ in $z$ direction. Again
the $g$-factors determined for the isolated atoms \cite{ari77} are
applied. The last line contains the spin-spin interaction
represented by the total molecular spin $S$ and its direction to
the molecule fixed axis $Z$. The parameter $\lambda$ is  a
function of $R$, mainly as $1/R^3$ from the magnetic dipole-dipole
interaction, but it contains also contributions from second order
spin-orbit interactions. The atomic masses for the kinetic energy
operator are taken from the recent tables by G. Audi et al
\cite{aud03}.

The calculations incorporate Born-Oppenheimer potentials  for
$U_X$ and $U_a$ according to Hund's coupling case b, because the
total electronic spin is taken as good quantum number. As start of
the analysis we apply directly the results of the Fourier
transform spectroscopy at Hannover \cite{pashov08} and simulate
the Zeeman pattern of the closest molecular levels just below
those asymptotes studied by the Feshbach resonances given in
Tab.\,\ref{TabI}. Through the crossings of these patterns as
function of magnetic field with the level of separated atoms we
obtain a first approximation of the expected Feshbach resonances.
These calculations were initially done for $^6$Li and $^{87}$Rb to
assign the results reported in \cite{Deh2008}. Because of the good
starting potentials it was almost straightforward to conclude that
the resonance at 1066.92\,G is a $s$-wave resonance and that at
882.02\,G is a $p$-wave resonance. This agrees to the result in
\cite{Li2008}. We made slight adjustments to the potentials at the
short range branches to bring the calculated Feshbach resonances
close to the observed positions. With these results predictions of
Feshbach resonances for the isotope combination  $^7$Li and
$^{87}$Rb were made and $s$-wave resonances were observed fairly
close to these predictions. During this study other resonances
were found which could be immediately assigned to be $p$-wave
resonances. Only the resonance around 535\,G could not be
identified up to now. This will be discussed later.

With this set of data a fit of the asymptotic form of the
potentials was performed, in which we fitted the long range
behavior with the van der Waals coefficient $C_6$ and $C_8$, to
bring the rotational barrier for the $p$-resonances to the right
position, and the short range branch. The fit identifies the
maximum of the rate constants of the corresponding elastic
two-body collisions at an average temperature of 4\,$\mu$K and
9\,$\mu$K for $^6$Li+$^{87}$Rb and $^7$Li+$^{87}$Rb
respectively as resonances. Because the spectroscopy analysis
\cite{pashov08} showed that for the singlet state a correction to
the Born-Oppenheimer potential has to be introduced to describe
the data set with $^6$Li and $^7$Li, this was also kept during the
long range analysis. For the triplet state a spectroscopic
data set from \cite{pashov08} is poor and thus in that analysis we
did not get a clear answer if a correction term would be needed.

\begin{figure}[ht]
\centerline{\scalebox{.8}{\includegraphics{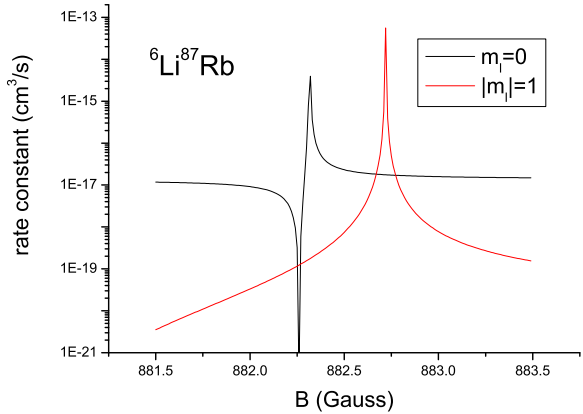}}}
\caption{(Color online)Splitting of p-wave resonance around 882\,G
for $^6$Li+$^{87}$Rb at a collision energy of 4\,$\mu$K, predicted
from pure spin-spin interaction.}
 \label{ml_split}
\end{figure}

Because the splitting of the $p$-wave resonances \cite{tic2004}
between the possible projections of the mechanical rotation
$m_l=0$ and $m_l=\pm 1$ onto the space fixed axis was not
observed, we included in the fit each $p$-resonance two times
assigning $m_l=0$ and +1 to both cases ($m_l=-1$ is coinciding
with +1)  and increased their uncertainty to obtain the proper
weighting of these observations for the fit. If one knew this
splitting one could get a direct measure of the spin-spin coupling
according to eq.\,(\ref{eq:ham}). We used the theoretical function
$\lambda (R)$ assuming pure spin-spin coupling of free electron
spins of the atoms. Fig. \ref{ml_split} shows the expected
splitting for the case $^6$Li+$^{87}$Rb, the rates were
calculated for a collision temperature of 4\,$\mu$K. The splitting
of about 0.5\,G is fairly large but was not observed in the
present experiment and also no asymmetry of the profile. These two
aspects probably indicate that the contribution by the second
order spin-orbit interaction compensates the pure spin-spin
interaction to a large extent, as it was already found for Rb$_2$.
We checked that thermal averaging broadens both resonances only
slightly, resolving them could be possible in future experiments
if two-body effects determine the resonance profiles.

\begin{table*}[ht]\tabcolsep2.mm
        \caption{Calculated Feshbach resonances for the two models in both the Fermi-Bose mixture $^6$Li+$^{87}$Rb
        and the Bose-Bose mixture $^7$Li+$^{87}$Rb. We plot the resonance positions $B_0$, and for the $s$-wave resonances
        the distances from the resonance positions to the lower lying zero crossings of the scattering length $\Delta_0^\infty$
        as well as the widths $\Delta B$ from fits to dispersion functions. In case of model I, the resonance position
        $B_{0,\text{rate}}$ refers to the maximum of the elastic scattering rate at finite temperature while $B_{0,\text{fit}}$
        is the resonance position for the dispersion fit at zero temperature. For the overlapping resonances of $^7$Li+$^{87}$Rb, the formula
        $a(B)=a_{\text{bg}}\,\left(1+\Delta B_1/(B-B_{\text{0,fit,1}})+\Delta B_2/(B-B_{\text{0,fit,2}})\right)$ was used. The fit interval
        and $a_{\text{bg}}$ are indicated in the corresponding column. The quantum numbers of the coupling molecular state
        are given in the last column.}\vspace{2mm}\renewcommand{\arraystretch}{1.5}
        \begin{tabular}[c]{c|cccc|ccc|c}\hline\hline
            \multirow{2}{*}{open channel} & \multicolumn{4}{c|}{model I}  & \multicolumn{3}{c|}{model II}&   assignment \\
 & $B_{0,\mathrm{rate}}\,(\text{G})$ & $B_{0,\mathrm{fit}}\,(\text{G})$ &$\Delta_0^\infty \,(\text{G})$&  $\Delta B\,(\text{G})$ & $B_{0,\mathrm{fit}}\,(\text{G})$ &$\Delta _0^\infty\,(\text{G})$&  $\Delta B\,(\text{G})$ & $\tilde{v},G,
 f,m_f,l$\\\hline
             \multirow{4}{*}{$^6$Li\,$|\frac{1}{2},\frac{1}{2}\rangle$+$^{87}$Rb\,$|1,1\rangle$} & 882.42 & &        && 882.04& & &-1, $\frac{3}{2}$, $\frac{5}{2}$, $\frac{3}{2}$, 1  \\
                                             & 1066.54 & 1066.51&-6.5 &7.4& 1066.93&-10.5& 10.5&-1, $\frac{3}{2}$, $\frac{5}{2}$, $\frac{3}{2}$, 0  \\
                                             &         & {\scriptsize $a_{\text{bg}}$=-14.3\,$a_B$}&& & {\scriptsize $a_{\text{bg}}$=-17.7\,$a_B$}& &  & \\
                                             &         & {\scriptsize $1000..1100$\,G}& & & {\scriptsize $1010..1140$\,G}&&  & \\\hline
            \multirow{6}{*}{$^7$Li\,$|1,1\rangle$+$^{87}$Rb\,$|1,1\rangle$} & 390.2 & &      & &388.9& & &-1, $\frac{3}{2}$, 3, 2, 1    \\
                                             & 445.6 & &     & &446.2& & &-1, $\frac{3}{2}$, 2, 2, 1    \\
                                             & 568.8 & 568.6 &-120 & 7.9&570.3& -115&7.4&-1, $\frac{3}{2}$, 3, 2, 0    \\
                                             & 651.6 & 650.7&-73  & 175&663.0& -87& 194.23&-1, $\frac{3}{2}$, 2, 2, 0  \\
                                             &         & {\scriptsize $a_{\text{bg}}$=-36\,$a_B$}& && {\scriptsize $a_{\text{bg}}$=-43\,$a_B$}& &  & \\
                                             &         & {\scriptsize $400..800$\,G}& & &{\scriptsize $400..800$\,G} &  & &\\\hline
            & $\sigma$=2.27  &  & & & $\sigma$=2.64 &  &\\\hline\hline

        \end{tabular}
        \label{TabII}
    \end{table*}

Table \ref{TabII} shows in column labeled "model I" the results of
this fit, where the resonances are described by the magnetic field
value $B_{\text{0,rate}}$ of the maximum of the two-body rate
constants and in cases of $s$-resonances the difference
$\Delta_0^\infty$ for the field where the rate goes to zero. We
also fit dispersion functions, in the case of the two overlapping
resonances of $^7$Li+$^{87}$Rb of the form
$a(B)=a_{\text{bg}}\,\left(1+\Delta
B_1/(B-B_{\text{0,fit,1}})+\Delta
B_2/(B-B_{\text{0,fit,2}})\right)$, to obtain the widths of the
individual resonances. The latter method is strictly only valid
for $T=0\,$K whereas the former gives the values in between which
the scattering length has its full dynamical range. Neither of the
two reported widths is equivalent to the width of the loss
features reported in Table \ref{TabI}. The assignment is described
by the vibrational quantum number $\tilde{v}$, counted from the
asymptote, and angular momenta with quantum number $G$, from the
coupling of the electronic spin with the nuclear spin of Rb
$\vec{G}=\vec{S}+\vec{i}_{\text{Rb}}$, the quantum number $f$ of
the total atomic angular momentum and its projection of the space
fixed axis $m_l$ and the mechanical rotation \textit{l}. Because
the hyperfine structure of Rb is much larger than that of Li, $G$
is in most cases a good quantum number, whereas the separated
atomic angular momenta loose their meaning. For the level just
below the asymptote both assignments could have comparable
quality, we choose the quantum number $G$ which becomes certainly
the better one when going to more deeply bound states. The
assignment shows also directly that the $p$-resonances correspond
to the same hyperfine levels as the $s$-resonances. Thus the
difference of the magnetic field between both groups reflect the
Zeeman energy which is about equal to the rotational energy. The
vibrational assignment $\tilde{v} =-1$ corresponds to the
conventional vibrational states counted from the bottom $v_a= 13$
and $v_X= 48$ for the mixed singlet and triplet states,
respectively, for $^6$Li+$^{87}$Rb and 14 or 52 for the case
$^7$Li+$^{87}$Rb. The obtained values for the long range potential
function, namely $C_6=2550.0\,$au and $C_8=2.3416\times10^5\,$au
are almost equal to the ones reported by theoretical estimates in
\cite{Derevianko01a,Porsev03} $C_6=2545(7)$\,au and
$C_8=2.34(4)\times10^5$\,au.

Comparing the experimental resonance positions with the results of the fit we
get significant deviations, sometimes outside the range of
experimental uncertainty. But there is no obvious systematic
trend. The resonances were observed by trap losses thus shifts by
three-body effects can be a concern. Studies in this direction are
planned.

In Tab.\,\ref{TabII} a different fit labeled "model II" is
presented, which was obtained completely independently by the
group in Berlin, using as starting potentials results from
\textit{ab initio} calculations and fitting as above the long
range and short range behavior. This fit reproduces the resonances
of the isotope pair $^6$Li+$^{87}$Rb almost exactly but shows
larger deviations for the other isotope combination. In the last
line of Tab. \ref{TabII} the normalized standard deviations are
given for both models applying the same weights for the individual
observations, and both are statistically only slightly different.
The deviation of $\sigma$ from a healthy value of about 1 signals
the existing problems in the present theoretical models like the
assumption of two-body collisions for describing trap loss
features. The long range parameters for model II are
$C_6=2543.0$\,au and $C_8=2.2825\times10^5$\,au. If one considers
the correlation of these parameters expected from the data
analysis, the two sets derived in this work are probably not
really different.

We were unable to assign the resonance found at 535\,G for
$^7$Li+$^{87}$Rb. For the experimentally prepared atomic pairs
$^7$Li\,$|1,1\rangle+^{87}$Rb\,$|1,1\rangle$ we checked the case
of further resonances by $s$-wave and $p$-wave and by $s$-wave
coupled to bound states with $l=2$ through spin-spin interaction.
Nothing was found in the desired vicinity. Sharp $s$-resonances by
the $l=2$ states are expected at magnetic fields below 150\,G. For
the other cases all expected resonances of the lowest atomic
asymptote are observed in this work for fields below 1.2\,kG. Only
very sharp resonances for $p$-waves with the selection rule
$\Delta m_l=-\Delta m_f\neq 0$ in the neighborhood of the much
stronger $p$-resonances $\Delta m_l=\Delta m_f=0$ are predicted,
but they are still too far from the observed one, namely 535~G,
and are too weak to be detected with experimental conditions in
our present experiment, mainly the observation time on resonance
and the sweep speed. Thus the question arises if the theory allows
an overall different assignment. This is clearly not the case,
because for the LiRb molecule, in which the small mass of the Li
atom vibrates with respect to the heavy Rb atom close to the
center of mass, the vibrational spacing is large. The next level
is about 30\,GHz below the last one at the asymptote. Thus the
hyperfine structure in LiRb is small compared to this spacing and
Feshbach resonances originating for these levels will only appear
above 5\,kG.

We also considered several other atomic asymptotes assuming that
the preparation of the atomic states by the rf field ramps might
be not pure enough. $s$-wave resonances for projection $m_f=0$ of
the total angular momentum (with possible combinations:
$m_f(\text{Rb})=\pm 1$ or 0 and $m_f(\text{Li})=\mp 1$ or 0) are
expected for fields larger than 700\,G and for $m_f=1$
$(m_f(\text{Rb})=1$ or 0 and $m_f(\text{Li})=0$ or 1) larger than
600 G (see Tab \ref{predict} below). Thus, these predictions are
too far from the observed 535\,G to try a fit with such an
assumption. Only $p$-resonances of those collision channels come
close to the observed value 535\,G, but then we have to accept two
weak points: first, the impure preparation of the collision pair
and second, the weakness of $p$-resonances. We might speculate
that this resonance does not originate from a two-body collision.
We already tried to study the collision pair
$^7$Li$\,\left|1,0\right\rangle+^{87}$Rb$\,\left|1,1\right\rangle$,
but the stability of such prepared mixture was too short, less
than 100\,ms, to allow a search for Feshbach resonances. The
collision pair
$^7$Li$\,\left|2,2\right\rangle+^{87}$Rb$\,\left|1,1\right\rangle$
could be prepared with good stability and we did not find a
resonance in the range of 12 to 1095\,G, which agrees to the
expectations from our theoretical results.

\begin{figure}[ht]
\centerline{\scalebox{.8}{\includegraphics{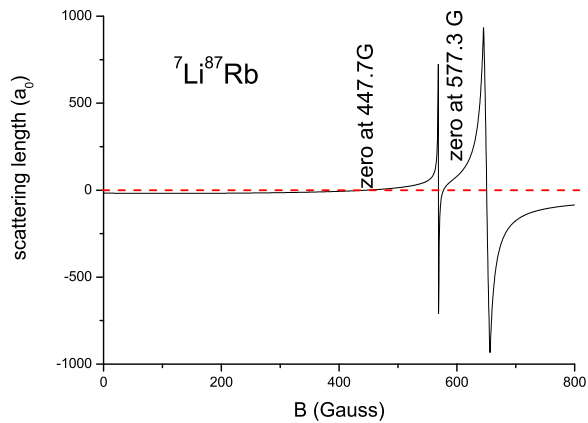}}}
\caption{(Color online) Scattering length of $s$-wave resonances
calculated for
$^7$Li$\,\left|1,1\right\rangle+^{87}$Rb$\,\left|1,1\right\rangle$
as function of magnetic field.}
 \label{prof} \end{figure}

For future experiments the tuning of interspecies interactions by
the two $s$-wave resonances in $^7$Li$^{87}$Rb is very promising.
Fig.\,\ref{prof} shows the calculated scattering length as
function of magnetic field. At low magnetic fields the scattering
length is very small and negative until the crossing around 438\,G
and is increasing slowly to the resonance at 566\,G. The long left
tail of the upper resonance almost overlaps with the lower one
thus the zero crossing between both is very close to the lower.
From the fit result in Tab. \ref{TabII} one sees, as mentioned
already earlier, that the observations are not completely
reproduced for these resonances, thus additional measurements
especially for the zero crossing would be desirable for preparing
the applicability of such interaction tuning. But independent of
this fact the convenient tunability of the interspecies
interaction with the help of these resonances remains valid.

With the new analysis we computed the scattering length of the
uncoupled singlet and triplet states for the different isotopic
combinations. The results are given in Tab.\,\ref{TabIII} with an
estimation of uncertainty limits (in units of the last digit shown
for the scattering length) for the observed isotope combinations
with $^{87}$Rb using the variability of the long range behavior
obtained during the fit series due to correlations between C$_6$
and C$_8$ considering both models. The scattering length of the
singlet state for $^6$Li+$^{87}$Rb is almost zero, thus the sign
is not yet determined, model I gives negative values whereas model
II would favor values with the opposite sign.

    \begin{table}[h]\tabcolsep1.5mm
        \caption{Calculated scattering lengths for the uncoupled singlet and triplet states for different isotopic combinations of Li and Rb
in Bohr radii (a$_0$ = 0.529$\times10^{-10}$m).}\vspace{2mm}
\renewcommand{\arraystretch}{1.5}
        \begin{tabular}[c]{cccccc}\hline\hline
            state          & model& $^6$Li$^{85}$Rb & $^6$Li$^{87}$Rb& $^7$Li$^{85}$Rb & $^7$Li$^{87}$Rb \\\hline
            \multirow{2}{*}{X$^1\Sigma^{+}$}&  I &   7.6 &  0.5(30)   &  60.5 &  53.9(20)  \\
                                            & II & 10.46 & 2.25       & 61.94 &  54.92     \\
            \multirow{2}{*}{a$^3\Sigma^{+}$}&  I & -14.3 &  -18.6(30) & -51.5 & -63.5(50)  \\
                                            & II & -15.50&  -19.81    & -55.08& -67.82     \\\hline\hline
        \end{tabular}
        \label{TabIII}
    \end{table}

The values for the triplet case are consistent with absolute
values determined by thermalization experiments on
$^6$Li+$^{87}$Rb \cite{Silber2005} and $^7$Li+$^{87}$Rb
\cite{Marzok2007}. The value of the triplet scattering length for
$^7$Li+$^{87}$Rb from the work by Li et al \cite{Li2008} is
inconsistent with the present result; this is not surprising
because these authors used constructed potentials for which they
did not know the true depth and thus the number of vibrational
levels each potential would accommodate. It is somehow surprising
that all triplet scattering lengths are small but have negative
values, thus attractive interspecies interaction will determine
ultracold ensembles at low magnetic fields.

\bigskip

In conclusion, we detected five collision resonances in a two
species ultracold mixture of $^7$Li and $^{87}$Rb gases with both
species in their respective absolute ground state
$\left|1,1\right>$. Four of them could be unambiguously assigned
to $s$- or $p$-Feshbach resonances. We combined these new data
with the existing observations on $^6$Li and $^{87}$Rb
\cite{Deh2008} to derive until now only unsatisfactorily known
interspecies interaction potentials of the Bose-Bose and
Fermi-Bose mixtures. For experiments employing the broad resonance
at 649\,G, low densities have to be used in order to decrease the
large losses associated with this resonance and to obtain a better
profile for determining the position of this resonance with higher
accuracy.

Because of the slightly different theoretical results we
calculated Feshbach resonances for other atomic collision
channels, for which further measurements could help to obtain a
converging analysis. Table \ref{predict} shows predictions for the
pair $^7$Li+$^{87}$Rb for some of the lowest atomic
asymptotes. Searching for the resonances of channel
$^7$Li\,$\left|1,0\right\rangle$+$^{87}$Rb\,$\left|1,1\right\rangle$
one should prepare the atomic pair at fields higher than 18\,G,
because at lower fields the losses to channel
$^7$Li\,$\left|1,1\right\rangle$+$^{87}$Rb\,$\left|1,0\right\rangle$
will be significant and for the case
$^7$Li\,$\left|1,-1\right\rangle$+$^{87}$Rb\,$\left|1,1\right\rangle$
the preparation should be above 174\,G to avoid inelastic losses
to
$^7$Li\,$\left|1,1\right\rangle$+$^{87}$Rb\,$\left|1,-1\right\rangle$
and
$^7$Li\,$\left|1,0\right\rangle$+$^{87}$Rb\,$\left|1,0\right\rangle$.
The predictions of the two models largely agree, but show a significant
difference for the Feshbach resonance in the channel
$^7$Li\,$\left|1,1\right\rangle$+$^{87}$Rb\,$\left|1,0\right\rangle$.
Studying this resonance could thus facilitate the discrimination between the
two models or their convergence. But this might be hampered by the fact that the resonance will
also be broadened by inelastic processes.

For the pair $^6$Li+$^{87}$Rb calculations were already
done by Li et al \cite{Li2008} giving few $s$-wave resonances
around 1000\,G or above, thus not in a convenient region for
experiments. Our own calculations show only small shifts for these
predictions, e.g. for the entrance channel
$^6$Li\,$\left|1/2,1/2\right\rangle$+$^{87}$Rb\,$\left|1,1\right\rangle$
we get 1292\,G instead of the reported value of 1278\,G in
\cite{Li2008}, no additional listing of these cases seems to be
necessary at present.

    \begin{table}[ht]\tabcolsep3mm
        \caption{Calculated Feshbach resonances for different collision channels in $^7$Li + $^{87}$Rb, positions given in
        G.}\vspace{2mm}\renewcommand{\arraystretch}{1.5}
        \begin{tabular}[c]{cccc}\hline\hline
            atomic channels          &\multirow{2}{*}{model I}& \multirow{2}{*}{model II} \\
            $^7$Li\,$\left|f,m_f\right\rangle$+$^{87}$Rb\,$\left|f,m_f\right\rangle$\\\hline
             & 634  & 644   \\
              $\left|1,0\right\rangle+\left|1,1\right\rangle$ & 663 &  667       \\
               & 748 & 763  \\\hline\vspace{2mm}
            $\left|1,1\right\rangle+\left|1,0\right\rangle$ & 786   &  807\rule[2mm]{0mm}{4mm} \\\hline
             & 717   &  730    \\
            $\left|1,-1\right\rangle+\left|1,1\right\rangle$   & 772 &  773       \\
               & 854 & 865      \\\hline\hline\vspace{1mm}
        \end{tabular}
        \label{predict}
    \end{table}
\bigskip

Future research could include formation of heteronuclear molecules
as mentioned in the introduction. For the population of highly
vibrationally excited states, such molecules could be formed by
either adiabatically sweeping the magnetic field over an
interspecies Feshbach resonance from high to low fields
\cite{Papp2006}, or by radio-frequency association
\cite{Ospelkaus2006}. In the case of the very desirable low lying
states, preferably the absolute ground state, Feshbach-assisted
photoassociation schemes \cite{vanAbeelen1998,Pellegrini2008}
could be the method of choice. The molecular spectroscopy started
by Pashov et al \cite{pashov08} will lead to the needed knowledge
of the rovibrational structure of electronically excited states.
Advancing to three-body effects, the scaling factor for Efimov
resonances \cite{Kraemer2006} is dependent on the mass ratio of
the collision partners \cite{Braaten2006}. For Li-Rb a much
smaller scaling factor is predicted than for three identical
bosons, making this system particularly interesting for possibly
uncovering universal properties of Efimov resonances. Further, by
fine-tuning the interspecies interaction in the Bose-Bose mixture,
mean-field induced stabilization of intrinsically unstable $^7$Li
condensates could be studied \cite{Marzok2007}.

\bigskip

This work has been supported by the Deutsche Forschungsgemeinschaft (DFG) through priority program SPP1116.

\end{document}